\documentclass[12pt,notitlepage]{article}
\usepackage{eucal,amsmath,amsfonts,amsthm,amssymb,latexsym}

\theoremstyle{plain}

\newtheorem{thm}{Theorem}

\newtheorem{Lma}{Lemma}

\newtheorem{PL}{PL}
\theoremstyle{definition}

\newcommand{\RR}{\mathbb{R}}

\newcommand{\rd}{\mathrm{d}}
\newcommand{\rw}{\mathrm{w}}

\newcommand{\ww}{\mathsf{\omega}}

\newcommand{\lp}{\overline{\lim}}
\newcommand{\lf}{\underline{\lim}}

\newcommand{\lindent}
{
\setlength{\labelwidth}{2cm}\setlength{\leftmargin}{2.0cm}
\setlength{\labelsep}{0.5cm}\setlength{\rightmargin}{1.0cm}
\setlength{\parsep}{0.5ex plus 0.2ex minus0.1ex}
\setlength{\itemsep}{0ex plus 0.2ex}
}

\newcommand{\findent}
{
\setlength{\labelwidth}{2cm}\setlength{\leftmargin}{1.7cm}
\setlength{\labelsep}{0.5cm}\setlength{\rightmargin}{1.0cm}
\setlength{\parsep}{0.5ex plus 0.2ex minus0.1ex}
\setlength{\itemsep}{0ex plus 0.2ex}
}

\newcounter{fact}
\newcounter{prop}

\newcommand{\ms}{{\setcounter{equation}{0}}{\setcounter{Lma}{0}}{\setcounter{Crly}{0}}\section}
\begin{document}
\setcounter{page}{1}

\title{Existence of Noncompact Static Spherically Symmetric Solutions of Einstein SU(2) Yang Mills Equations \thanks{Part of a doctoral thesis at the University of Michgan under the supervision of Professor Joel A. Smoller}}
\author{Alexander N. Linden\thanks{Zorn Assistant Professor, Indiana University}}
\date{}
\maketitle


\begin{abstract}
We consider static spherically symmetric solutions of the Einstein equations with cosmological constant $\Lambda$ coupled to the SU(2) Yang Mills equations that are smooth at the origin $r=0$.  We  prove that all such solutions have a radius $r_c$ at which there is a horizon.  However, for any positive integer $N$, there exists a small positive $\Lambda_N$, such that whenever $0<\Lambda<\Lambda_N$, there exist at least $N$ distinct solutions for which the singularity is only a coordinate singularity and the solution can be extended to $r>r_c$. 

\end{abstract}
\ms{Background}
\label{intro}
\subsection{Introduction}
$\;$

There have been numerous studies of the Einstein equations without cosmological constant coupled to the SU(2)-Yang Mills equations ever since Bartnik and McKinnon found numerical evidence of (particlelike) solutions that are globally smooth and asymptotically  Minkowski space.  Subsequently, Smoller and Wasserman established rigorously, the existence of an infinite family of such solutions (\cite{jS93}).  In this paper, we consider the same system, but with a small positive cosmological constant $\Lambda$.  We prove, in Theorem~\ref{Acrashes}, that no such solution can be regular in the sense of the particlelike solutions.  Specifically, we prove that every solution that is regular at the origin must give rise to an horizon.  In Theorem~\ref{ppl} we prove the existence of solutions in which this horizon is only a coordinate singularity; i.e., solutions in which the apparent singularity is due only to choice of Schwarzschild coordinates.  Such solutions, (which we call noncompact), can be identified by the number of zeros of one of the coefficients of the Yang-Mills connection.\\

The static spherically symmetric Einstein-Yang-Mills equations with the cosmological constant take the form of two differential equations for the variables $A(r)$ and
$\rw(r)$:
\begin{equation}\label{Aeq}
rA'+2A{\rw'}^2=\Phi\;\mathrm{and}
\end{equation}
\begin{equation}\label{weq}
r^2A\rw''+r\Phi\rw'+\rw(1-\rw^2)=0
\end{equation}
where 
\begin{equation}\label{phidef}
\Phi=1-A-\frac{(1-\rw^2)^2}{r^2}-\Lambda r^2.
\end{equation}
$A$ is the same $A$ that appears in a spherically symmetric metric written as
\begin{equation}\label{metric}
\rd s^2 = -C^2(r,t)A(r,t)\; \rd t^2 + \frac{1}{A(r,t)} \;\rd r^2
+\rd \Omega^2
\end{equation}
\[(\rd\Omega^2=r^2 \;\rd \phi^2+r^2\sin^2\phi\; \rd \theta^2)\]
and w is the same w that appears in the spherically symmetric connection on an
\textbf{SU}(2) bundle; namely,
\begin{eqnarray}\label{connection}
\ww&=& \mathrm{a}(r,t) \mathbf{\tau}_3 \;\rd t + \mathrm{b}(r,t)\mathbf{\tau}_3 \;\rd r +\rw(r,t)\mathbf{\tau}_2
\;\rd \phi\nonumber\\
&+&(\cos\phi\mathbf{\tau}_3-\rw(r,t) \sin \phi
\mathbf{\tau}_1)\;\rd \theta.
\end{eqnarray}
$\mathbf{\tau}_i$ are the following matrices which form a basis of $su(2)$:
\[\mathbf{\tau}_1=i/2\left[
\begin{array}{cc}
0 & -1\\
-1 & 0
\end{array}
\right],\;
\mathbf{\tau}_2=i/2\left[
\begin{array}{cc}
0 & i\\
-i & 0
\end{array}
\right],\;
\mathbf{\tau}_3=i/2\left[
\begin{array}{cc}
-1 & 0\\
 0 & 1
\end{array}
\right].
\]
There is also an equation for $C$,
\begin{equation}\label{Ceq}
rC'=2{\rw'}^2C.
\end{equation}
However, equation~(\ref{Ceq}) separates from equations~(\ref{Aeq}) and (\ref{weq}) and yields
\begin{equation}\label{Csolved}
C(r)=C(r=0)e^{\int_{t=0}^r(2{\rw'}^2/t)\;\rd t}
\end{equation}
where $C(r=0)$ can be assigned arbitrarily.

Throughout this paper, we will make use of the following equation (for different choices of $\beta$) that is easily obtained from equations~(\ref{Aeq}) and (\ref{weq}):
\begin{equation}\label{Aw'beta}
r^2(A{\rw'}^\beta)'+r{\rw'}^\beta[(\beta-1)\Phi+2A{\rw'}^2]+\beta\rw{\rw'}^{\beta-1}(1-\rw^2)=0.
\end{equation}
Particularly useful are the cases $\beta=1$,
\begin{equation}\label{veq}
r^2(A{\rw'})'+2r{\rw'}^2(A\rw')+\rw(1-\rw^2)=0
\end{equation}
and $\beta=2$,
\begin{equation}\label{feq}
r^2(A{\rw'}^2)'+r{\rw'}^2[\Phi+2A{\rw'}^2]+2\rw\rw'(1-\rw^2)=0.
\end{equation}

We begin by stating some simple facts and preliminary results.  In Section~\ref{Pseudo}, we prove Theorems~\ref{Acrashes} and \ref{ppl}.  Section~\ref{tech} contains proofs of the technical lemmas used in Chapter~\ref{Pseudo}.     
\subsection{Preliminaries}
$\;$

In this section, we state some basic facts regarding solutions to Equations~(\ref{Aeq}) and (\ref{weq}).
\begin{list}{\textit Fact \arabic{fact}:}
{\usecounter{fact}
\findent}
\item \label{reflect}The equations are invariant under the transformation $(r,A,\rw)\rightarrow (r,A,-\rw)$.
\end{list}
\begin{list}{\textit Fact \arabic{fact}:}
{\usecounter{fact}
\addtocounter{fact}{1}
\findent}
\item \label{wconst} If $\rw$ is constant, equation~(\ref{weq}) implies $\rw\equiv \pm 1$ or $\rw\equiv 0$.
\end{list}
Integrating equation~(\ref{Aeq}) with $\rw^2\equiv 1$ yields
\begin{equation}\label{AdeSit}
A=1-\frac{2M}{r}-\frac{\Lambda r^2}{3}
\end{equation}
where $M$ is an arbitrary constant.  With a possibly rescaled $t$, this is a deSitter space with constant Yang Mills connection,
\begin{eqnarray}\label{deSitter}
\rd s^2&=&-(1-\frac{2M}{r}-\frac{\Lambda r^2}{3})\rd t^2+(1-\frac{2M}{r}-\frac{\Lambda r^2}{3})^{-1}\rd r^2+r^2\rd \Omega ^2\nonumber\\
\rw^2&\equiv& 1.
\end{eqnarray}
If $\rw\equiv 0$, another simple calculation yields
\begin{equation}\label{RN}
A(r)=1-\frac{\Lambda r^2}{3}+\frac{1}{r^2}+\frac{c}{r};\;\;\;\rw\equiv 0
\end{equation}
where $c$ is an arbitrary constant.
If, for any $r>0$, ${\rw'}(r)=0$ and ${\rw}^2(r)=1$ or $0$, then by uniqueness of solutions of ordinary differential equations, the solution must be~(\ref{deSitter}) or~(\ref{RN}).
\begin{list}{\textit Fact \arabic{fact}:}
{\usecounter{fact}
\addtocounter{fact}{2}
\findent}
\item\label{Einstsp}  There is another known explicit solution; namely Einstein space, when $\Lambda=3/4$, $A=1-r^2/2$, and $\rw=\sqrt{A}$.
\end{list}
\begin{list}{\textit Fact \arabic{fact}:}
{\usecounter{fact}
\addtocounter{fact}{3}
\findent}
\item \label{family} Given any $\Lambda$ and $\lambda$ there is an interval $I_\lambda=[0,r_\lambda)$ in which there exists a solution of
equations (\ref{Aeq}) and (\ref{weq}) with the following properties:
\begin{list}{(\Roman{prop}):}
{\usecounter{prop}
\lindent}
\item $\lim_{r\searrow 0}(A(r),\rw^2(r),\rw'(r),\rw''(r))=(1,1,0,-\lambda)$,
\item The solution is analytic in the interior of $I_\lambda$,
and $\mathcal{C}^{2+\alpha}$ in $I_\lambda$ for a small $\alpha>0$.
\item The solutions depend continuously on $\lambda$.
\end{list}
\end{list}
A proof of \textit{Fact~\ref{family}} in the case $\Lambda = 0$ can be found in \cite{js91}.  The same proof is valid with minor modification in the general case $\Lambda\ne 0$.\\

We define the region
\[\Gamma=\{(r,A,\rw,{\rw'}):r>0,\;A>0,\; {\rw}^2\le 1\; \mathrm{and} \;(\rw,{\rw'})\not=(0,0)\}.\]
We then set $r_c$ to be the smallest value of $r$ that satisfies $A(r_c)=0$ if such an $r$ exists and set $r_c=\infty$ if no such $r$ exists.
\pagebreak
$\;$\\
\begin{center}
Figure~1.
\end{center}
\setlength{\unitlength}{0.7mm}
\begin{picture}(0,100)(-40,0)\label{pexit}
\put (10,15){\vector(0,1){80}}
\put (0,50) {\vector(1,0){100}}
\put (10,70) {\line(1,0){90}}
\put (3.5,45){$(0,0)$}
\put (101,49){$r$}
\put (8,95.5){w}
\put (5,68){$1$}
\qbezier (20,20) (40,130)(80,55)
\put (30,50){\dashbox(0,10)}
\put (27.1,45){$r_0$}
\put (34.3,50){\dashbox(0,20)}
\put (33.4,45){$\bar\rho$}
\put (73,50){\dashbox(0,17.1)}
\put (72,45){$r_1$}
\put (60,50){\dashbox(0,32)}
\put (58,45){$r_e$}
\put (80,50){\dashbox(0,5)}
\put (80,45){$r_c$}
\put (25,20){$\rw(\bar r)>1,\;\rw'(\bar r)=0$, $\rw''(\bar r)\le 0$}
\put (51,50){\dashbox(0,35.45)}
\put (49.5,45){$\bar r$}
\end{picture}
\par
\indent
\nopagebreak
\begin{list}{\textit Fact \arabic{fact}:}
{\usecounter{fact}
\addtocounter{fact}{4}
\findent}
\item\label{permexit} Suppose for some $r_0\ge 0$, $(r_0,A(r_0),\rw(r_0),\rw'(r_0))\in\Gamma$ but $(r_e,A(r_e),\rw(r_e),\rw'(r_e))\notin \Gamma$ for some $r_e\in(r_0,r_c)$.  Then
$(r,A(r),\rw(r),\rw'(r))\notin \Gamma$ for all $r\in [r_e,r_c)$.
\end{list}
\textbf{Proof of Fact~\ref{permexit}}:  At $r_e$, one of the following must hold:
\newcounter{whenexit}
\begin{list}{(\arabic{whenexit}):}
{\usecounter{whenexit}
\lindent}
\item\label{wea} $A(r_e)=0$,
\item\label{web} $(\rw(r_e),\rw'(r_e))=(0,0)$, or
\item\label{wec} ${\rw(r_e)}^2>1$.
\end{list}
We now examine each of these cases.\\ \\
\textit{Case~\ref{wea}}.  $r_e=r_c$ and so there is nothing to prove.\\ \\
\textit{Case~\ref{web}}.  From \textit{Fact}~\ref{wconst}, $\rw\equiv 0$, contrary to the hypotheses.\\ \\
\textit{Case~\ref{wec}}.  Because the equations are smooth in the region $[r_0,r_1]$, $\rw$ is also smooth in this region.  It follows that there exists a $\bar\rho\in (r_0,r_1)$ that satisfies ${\rw}^2(\bar
\rho)=1$ and ${\rw'}\rw(\bar \rho)\ge 0$.  If ${\rw'}(\bar \rho)=0$, then by \textit{Fact}~2, ${\rw}^2\equiv 1$.  We may therefore assume that ${\rw'}(\bar \rho)\ne 0$.  Thus, there exists also an $\bar r\in (\bar \rho,r)$ such that
${\rw'}(\bar r)=0$ and $\rw''\rw(r)<0$; (see Figure~1).  But equation~(\ref{weq}) shows that this is impossible.\hfill $\blacksquare$
\begin{list}{\textit Fact \arabic{fact}:}
{\usecounter{fact}
\addtocounter{fact}{5}
\findent}
\item\label{wblowup} Let $r_0$ and $r_c$ be such that $r_0<r_c$, $(r_0,A(r_0),\rw(r_0),\rw'(r_0))\in\Gamma$ and $\lim_{r\nearrow r_c}A(r_c)=0$.  If there exists a $\rho$ in $(r_0,r_c)$ that satisfies $(\rho,A(\rho),\rw(\rho),{\rw'}(\rho))\notin \Gamma$, then $\lim_{r\nearrow r_c}{\rw'}^2(r)=\infty$.
\end{list}
\textbf{Proof of Fact 6.}  The Intermediate Value Theorem guarantees the existence of an $r_e\in(r_0,\rho)$ such that ${\rw}^2(r_e)=1$.  \textit{Fact}~\ref{permexit} states that ${\rw}^2(r)>1$ for all $r\in(r_e,r_c)$.  Using \textit{Fact}~\ref{reflect}, we lose no generality by assuming $\rw(r_e)=1$.  \textit{Fact}~\ref{wconst} ensures that $\rw'(r_e)>0$.  As in the proof of \textit{Fact}~\ref{permexit}, equation~(\ref{weq}) implies that $\rw'(r)>0$ in the entire interval $(r_e,r_c)$; i.e., $\lim_{r\nearrow r_c}\rw(r)$ exists.

We now prove $\lim_{r\nearrow r_c}\rw'(r)$ exists also. To this end, it suffices to find a $\rw'_c$, possibly infinite, such that $\lim_{r_n\nearrow r_c}\rw'(r_n)=\rw'_c$ for any sequence $\{r_n\}$ that approaches $r_c$ and also satisfies $\rw''(r_n)=0$.  On any sequence, $\{r_n\}\nearrow r_c$ for which $\rw''(r_n)=0$, 
\begin{equation}\label{wpseq} 
\rw'(r_n)=-\frac{\rw(r_n)(1-{\rw}^2(r_n))}{r_n\Phi(r_n)}.
\end{equation}
Since $\rw(1-{\rw}^2)$ has a nonzero limit as $r\nearrow r_c$, it follows from equation~(\ref{phidef}) that $\Phi$ also has a limit as $r\nearrow r_c$.  Consequently, the right side of equation~(\ref{wpseq}) has a limit; i.e., $\lim_{r\nearrow r_c}\rw'(r)$ exists.

If $\lim_{r\nearrow r_c}\rw'(r)$ is finite, then $\lim_{r\nearrow r_c}A{\rw'}(r)=0$. Since $A\rw'(r)>0$  in a neighborhood of $r_c$, there exists a sequence $\{s_n\}\nearrow r_c$ such that 
\begin{equation}\label{vseq}
(A\rw')'(s_n)<0.
\end{equation}
However, equation~(\ref{veq}) implies that
\begin{equation}\label{vcontr}
(A\rw')'(r)>0\;\mathrm{for\;all}\;r\;\mathrm{near}\;r_c.
\end{equation}
Clearly both equation~(\ref{vseq}) and equation~(\ref{vcontr}) cannot hold.   We conclude $\lim_{r\nearrow r_c}\rw'(r)=\infty$.\hfill $\blacksquare$\\

The significance of \textit{Fact}~\ref{wblowup} is the following:
\begin{list}{\textit Fact \arabic{fact}:}
{\usecounter{fact}
\addtocounter{fact}{6}
\findent}
\item\label{krusk} There is a non-singular change of coordinates such that the metric takes the form
\[ \rd s^2 = f^2(u,v)(\rd v^2 -\rd u^2)+r^2(\rd \phi^2 +\sin^2\phi\; \rd \theta^2)\]
near $r_c$ whenever $\lp_{r\nearrow r_c}{\rw'}^2(r)<\infty$.  When this holds, the transformation 
is such that the Yang Mills curvature $F$ is nonsingular near $r_c$ and such that $r>r_c$ in the extended solution.
\end{list}
A proof of \textit{Fact}~\ref{krusk} can be found in \cite{js951}.  The proof is valid with minor modification also when $\Lambda\ne 0.$

\pagebreak
\ms{Existence of Noncompact Solutions}
\label{Pseudo}
$\;$

In this section, we establish the existence of noncompact solutions.  Such solutions are characterized by the property that they are smooth for $0\le r<r_c$ where $r_c$ is a finite value (crash point) at which $A=0$.   Although, as we shall see, all solutions which are smooth at the origin have a crash point, the noncompact solutions are distinguished by the fact that the singularity at $r_c$ can be removed by a Kruskal-like change of coordinates in which $r>r_c$ in the extended solution.  Furthermore, the Yang Mills field is well behaved under the change of coordinates.
\subsection{Outline of Existence Proof}
\label{outline}
$\;$

The existence proof is based on three theorems relating to particlelike solutions which hold only in the case  $\Lambda=0$.   

\begin{PL}\label{particle}
For each $n>0$ there is a solution $(A_n(r),\rw_n(r))$ of equations~(\ref{Aeq}) and (\ref{weq}) that satisfies
the following conditions:

\begin{list}{(\Roman{prop}):}
{\usecounter{prop}
\lindent}
\item$(A_n(0),\rw_n(0),\rw_n'(0))=(1,1,0)$,
\item the solution is regular (i.e.,$ A(r)>0$) for all $r>0$, 
\item$\lim_{r\nearrow \infty}(A_n(r),\rw_n^2(r),\rw_n'(r))=(1,1,0)$, and
\item$\rw_n$ has $n$ zeros.
\end{list}
\end{PL}
\noindent
\begin{PL}\label{54}
Let $\bar P_n=(\bar r_n,\bar A_n,\bar\rw_n,\bar \rw'_n)$ be a sequence of points in $\Gamma$ 
such that 
\[\bar \rw_n^2\rightarrow 1,\;\bar r_n\rightarrow \infty\; \mathrm{and}\;\bar r_n(1-\bar A_n)<M\]
for some $M>0$.  Let $P_n(r)=(r,A(r),\rw(r),\rw'(r),\rw'(r))$ be the orbit through $\bar P_n$, defined for
$r>r_n$ and suppose that
\[0\le \rw'(r_n)/\rw(r_n)\le 1.\]
Then for sufficiently large $n$, $P_n(r)$ exits $\Gamma$ through $\rw^2=1$, at $r=r_e^n$ and $\Theta(\bar
\rw,\bar \rw')-\Theta(\rw(r_e^n),\rw'(r_e^n))<5\pi/4$ where $\Theta(\rw,\rw')=\arctan(\rw'/\rw)$.\\ \\
\end{PL}
\noindent
\begin{PL}\label{lambdabar}
There exists a $\bar \lambda$ , $1<\bar \lambda<2$ such that whenever $\lambda<\bar \lambda$, $A>0$ as long as the
orbit is in $\Gamma$.
\end{PL}  
\noindent
\cite{jS93} contains a proof of PL~\ref{particle} and PL~\ref{54}.  A proof of
PL~\ref{lambdabar} can be found in \cite{jS94}.\\

\textit{Fact}~\ref{family} gives a continuous one parameter family of solutions that are smooth in a neighborhood
of $r=0$ and satisfy $A(0)=1$, $\rw(0)=1$, $\rw'(0)=0$ and
$\rw''(0)=-\lambda<0$.  Throughout the remainder of this paper, unless otherwise stated, all solutions $(A,\rw)$ are members of this family.  We will
consider $\Lambda$ and $\lambda$ as parameters and write solutions as 
\[(A(\Lambda,\lambda,r),\rw(\Lambda,\lambda,r))\]
when necesssary to avoid ambiguity.

Considering \textit{Fact}~\ref{krusk}, it suffices to find solutions for which $\rw'(r)$ remains finite as $r\nearrow r_c$.  We will show that there exist solutions for which $\rw'$ not only remains bounded but for which $\lim_{r\nearrow r_c}\rw'(r)$ exists.  Such solutions will be found by a perturbation argument that can be described as follows:

PL~\ref{54} 
gives solutions
$(A(0,\lambda_n,r),\rw(0,\lambda_n,r))$ that satisfy the following conditions:
\begin{list}{(\Roman{prop}):}
{\usecounter{prop}
\lindent}
\item $\rw(0,\lambda_n,r)$ has $n$ zeros,
\item the orbit of $(A(0,\lambda_n,r),\rw(0,\lambda_n,r))$ leaves $\Gamma$ through $\rw=-1$ if $n$ is odd, the orbit of $(A(0,\lambda_n,r),\rw(0,\lambda_n,r))$ leaves $\Gamma$ through $\rw=1$ if $n$ is even, and
\item $A(r)>0$ at the exit point $r_e$.
\end{list}
(See \cite{jS93}).

Perturbing these solutions by changing  $\Lambda$  will give similar solutions\\
 $(A(\Lambda,\lambda_n,r),\rw(\Lambda,\lambda_n,r))$ and $(A(\Lambda,\lambda_{n+1},r),\rw(\Lambda,\lambda_{n+1},r))$ 
provided $\Lambda$ is small.  We will fix such a small $\Lambda$ and consider all solutions
$(A(\Lambda,\lambda,r)$,
$\rw(\Lambda,\lambda,r))$ where $\lambda$ is between $\lambda_n$ and $\lambda_{n+1}$. 
One of the perturbed solutions
will be noncompact.

The perturbation argument requires that $A'$, $\rw$, and $\rw'$ be well behaved near $r_c$ and that $r_c$ be a continuous function of $\lambda$.  This may not be the case when $r_c^2\in J_\Lambda$ where $J_\Lambda$ is the interval defined by 
\begin{equation}\label{J}
J_\Lambda=(0,2]\cup[\frac{1-\sqrt{1-4\Lambda}}{2\Lambda},\frac{1}{\Lambda}].
\end{equation}
We will eliminate this possibility.
\subsection{Existence Proof}
\label{Existence}
$\;$

The next Theorem shows that the cosmological constant precludes globally regular solutions in the sense of PL~\ref{particle}.
\begin{thm}\label{Acrashes}
For any $\Lambda>0$ and $\lambda$, there exists an $r_c(\Lambda,\lambda)\le\sqrt{3/\Lambda}$ such that either $\lim_{r\nearrow r_c}A(r)=0$ or $\lp_{r\nearrow r_c}{\rw'}^2(r)=\infty$.
\end{thm}
\noindent
\textbf{Proof}:  We suppose a solution $(A,\rw)$ to be valid up to $r_c$ and consider the following function:
\begin{equation}\label{mu}
\mu(r)=r(1-A-\Lambda r^2/3).
\end{equation}
A simple calculation using equation~(\ref{Aeq}) yields
\begin{equation}\label{mu'}
\mu'=2A{\rw'}^2+\frac{(1-\rw^2)^2}{r^2}.
\end{equation}
Now, obviously, $\mu'\ge 0$ whenever $A>0$.  Moreover, $A>0$ whenever $\mu<0$ and $r<\sqrt{3/\Lambda}$.  If, for any $\hat r<\min\{r_c,\sqrt{3/\Lambda}\}$, $\mu(r)<0$, then $\mu(r)<0$ for all $r\in[0,\hat r]$.  In particular, $\mu(0)<0$ and, thus, $A(0)>1$.  Because we are considering only solutions in the family of \textit{Fact}~\ref{family}, we conclude that
\begin{equation}\label{atrap}
0<A(r)\le 1 -\Lambda r^2/3\;\;\;\mathrm{whenever}\;r\in(0,\min\{r_c,\sqrt{3/\Lambda}\}).
\end{equation}

If for some  $\hat r \in (0,\min\{r_c\sqrt{3/\Lambda}\})$, $\rw^2(\hat r)=1$ and $\rw'(\hat r)=0$, then \textit{Fact}~\ref{wconst} implies equations~(\ref{Aeq}) and (\ref{weq}) have a solution of the form
\[A(r)=1-\frac{\Lambda r^2}{3}+(A(\hat r)+\frac{\Lambda \hat r^2}{3}-1)\frac{\hat r}{r}.\]
The only such solution that satisfies $A(0)=1$ is 
\[A(r)=1-\Lambda r^2/3.\]
But this solution clearly satisfies the Theorem.

To complete the proof, we need consider only the remaining situation; that in which for all $ r \in (0,\min\{r_c,\sqrt{3/\Lambda}\})$, either 
$\rw^2(r)\ne 1$ or $\rw'(r)\ne
0$; i.e., $\mu'>0$.  In this case, there are $\epsilon>0$ and $\delta\in(0,\min\{r_c,\sqrt{3/\Lambda}\})$, such that $\mu(r)>\epsilon$ whenever $r\in(\delta,\min\{r_c,\sqrt{3/\Lambda}\})$.  On the other hand, if $r_c\ge 3/\sqrt{\Lambda}$, then equation~(\ref{mu}) shows that $\mu(\sqrt{3/\Lambda})<0$.  It must be, then, that $r_c<\sqrt{3/\Lambda}$.
\hfill $\blacksquare$\\ 

We also have the following:
\begin{Lma}\label{a=0}
$\lim_{r\nearrow r_c}A(r)$ exists.  Furthermore, if both $\lim_{r\nearrow r_c}A(r)>0$ and $\lp_{r\nearrow r_c}\rw^2(r)<\infty$, then $\lp_{r\nearrow r_c}{\rw'}^2(r)<\infty$.
\end{Lma}
\noindent
\textbf{Proof}:  Equation~(\ref{Aeq}) implies $A'$ is bounded from above in a neighborhood of $r_c$.  Thus, $\lim_{r\nearrow r_c}A(r)$ exists.  If $\lim_{r\nearrow r_c}A(r)>0$ and there exists a sequence $\{r_n\}\nearrow r_c$ that satisfies $\lim_{n\nearrow \infty}\rw'(r_n)=\infty$, then $\lim_{n\nearrow \infty}A\rw'(r_n)=\infty$ and the sequence can be chosen so that $(A\rw')'(r_n)>0$ also.  However, equation~(\ref{veq}) shows that there can be no such sequence.\hfill $\blacksquare$
$\;$\\

We now make precise the set of solutions subject to perturbation by defining
\[\mathcal{K}(\Lambda_0,\lambda_0)=\{(\Lambda,\lambda):0\le\Lambda\le \Lambda_0\,\mathrm{and}\;0\le\lambda\le\lambda_0\}.\]
In Section~\ref{tech} we will establish the existence of $\Lambda_0$ and $\lambda_0$ such that the set
\begin{equation}\label{Cset}
\mathcal{K}=\mathcal{K}(\Lambda_0,\lambda_0)
\end{equation}
satisfies the following conditions:
\begin{list}{(\Roman{prop}):}
{\usecounter{prop}
\lindent}
\item all solutions subject to perturbation lie in $\mathcal{K}$,
\item for all solutions in $\mathcal{K}$, $\lim_{r\nearrow r_c}(A'(r),\rw(r),\rw'(r))=(A'_c,\rw_c,\rw'_c)$ exists.  Moreover, both $A'_c$ and $\rw_c$ are finite, and 
\item for any $\Lambda\in (0,\Lambda_0]$, $r_c(\lambda)$ is a continuous function of $\lambda$.
\end{list}
For each solution $(A(\Lambda,\lambda,r),\rw(\Lambda,\lambda,r))$, $(\Lambda,\lambda)\in\mathcal{K}^o$ (interior of $\mathcal{K}$), there exists a smallest $r_c>0$ at which the solution leaves $\Gamma$. Given a $\mathcal{K}$, we define the subsets
\begin{equation}\label{C_c}
\mathcal{K}_c=\{(\Lambda,\lambda)\in \mathcal{K}^o:A(r_c)=0\}\; \mathrm{and}
\end{equation}
\begin{equation}\label{Cccomp}
\bar\mathcal{K}_c=\{(\Lambda,\lambda)\in \mathcal{K}^o:A(r_c)>0\}.
\end{equation}
We will refer to a solution $(A(\Lambda,\lambda,r),\rw(\Lambda,\lambda,r))$ as a \textbf{solution in}  $\mathbf{\mathcal{K}}$, $\mathbf{\mathcal{K}_c}$ or $\mathbf{\bar\mathcal{K}_c}$ whenever $(\Lambda,\lambda)$ is in any of these respective sets.

To distinguish $\mathcal{K}_c$ from $\bar\mathcal{K}_c$, we denote $r_c$ by $r_e$ for solutions in $\bar\mathcal{K}_c$; i.e.,  $r_e(\Lambda,\lambda)$ satisfies $\rw^2(r_e)=1$ and $A(r_e)>0$ for solutions that exit $\Gamma$ with positive $A$.

Lemma~\ref{a=0} precludes the possibility that both
\[\lp_{r\nearrow r_c}\rw^2(r)\le 1,\;\mathrm{and}\;\lf_{r\nearrow r_c}A(r)>0\]
and
\[\lp_{r\nearrow r_c}{\rw'}^2(r)=\infty\]
hold simultaneously; i.e., an orbit cannot leave $\Gamma$ because $\rw'$ blows up.\\

The continuity of $r_c$ as a function of $\lambda$ will be proved in the general case in Section~\ref{tech}.  Here, we prove that the set $\bar\mathcal{K}_c$ is open and $r_c(\lambda)$ restricted to $\bar\mathcal{K}_c$ is continuous.

\begin{Lma}\label{Arc+}
Suppose $(\bar\Lambda,\bar\lambda)\in \bar\mathcal{K}$ and $\rw(\bar\Lambda,\bar\lambda,r_e(\bar\Lambda,\bar\lambda))=\pm 1$.  Then for every $\epsilon>0$ there exist $\delta>0$ such that whenever $\max\{|\Lambda-\bar\Lambda|,|\lambda-\bar\lambda|\}<\delta$, $|r_e(\Lambda,\lambda)-r_e(\bar\Lambda,\bar\lambda)|<\epsilon$, $A(r_e(\Lambda,\lambda))>0$ and $\rw(\Lambda,\lambda,r_c(\Lambda,\lambda))=\pm 1$.
\end{Lma}
\noindent
\textbf{Proof}:  Lemma~\ref{a=0} implies that any solution $(A,\rw)$ that satisfies  $A(r_e)>0$, also satisfies $\lp_{r\nearrow r_e}{\rw'}^2(r)<\infty$.  From standard theorems it follows that the solution $(A(\bar\Lambda,\bar\lambda),\rw(\bar\Lambda\bar\lambda))$ can be extended to some $\tilde r>r_e(\bar\Lambda,\bar\lambda)$ with $A>0$ whenever $r<\tilde r$.  Also, invoking \textit{Fact}~\ref{reflect} and \textit{Fact}~\ref{wconst}, we may assume that $\rw(\bar\Lambda,\bar\lambda,r_e(\bar\Lambda,\bar\lambda))=1$ and $\rw'(\bar\Lambda,\bar\lambda,r_e(\bar\Lambda,\bar\lambda))>0$. There then exist $\eta>0$ and $r_\eta\in(r_e,\tilde r)$ such that
\[\rw(\bar\Lambda,\bar\lambda,r_\eta)=1+2\eta.\]
Continuous dependence on parameters and standard theorems guarantee the existence of a neighborhood $V=(\bar\Lambda-\delta,\bar\Lambda+\delta)\times(\bar\lambda-\delta,\bar\lambda+\delta)$ such that whenever $(\Lambda,\lambda)\in V$, the solution $A(\Lambda,\lambda,r)$ exists beyond $r_\eta$ and $\rw(\Lambda,\lambda,r_\eta)>1+\eta>1$; i.e., $(A,\rw)$ also exits $\Gamma$ at $r_e(\Lambda,\lambda)$ through $\rw=1$.

It remains to prove that $r_e(\Lambda,\lambda)$ is continuous.  
 $\rw(\Lambda,\lambda,r)$ and $\rw'(\Lambda,\lambda,r)$ are continuous in $r$ at $(\bar\Lambda,\bar\lambda,r_e(\bar\lambda,\bar\lambda))$ and $\rw'(\bar\Lambda,\bar\lambda,r_e(\bar\Lambda,\bar\lambda))\ne 0$. The Implicit Function Theorem gives the continuity of $r_c$ locally.  \textit{Fact}~\ref{permexit} implies that there is no other $r>0$ satisfying $\rw^2(r)=1$. This completes the proof.  \hfill $\blacksquare$\\ 

In the process of proving Lemma~\ref{Arc+} we also proved the following:
\begin{Lma}\label{A+limits}
For any $(\Lambda,\lambda)\in\bar\mathcal{K}_c$, $A'$ and $\rw'$ have finite limits as $r\nearrow r_e(\Lambda,\lambda)$.
\end{Lma}

The existence of noncompact solutions is a corollary of the following which assumes results to be proved in Section~\ref{tech}:
\begin{Lma}\label{wpinfopen}
Suppose ${\rw'}(\Lambda,\lambda_0,r_c(\Lambda,\lambda_0))=\pm\infty$.  Then there exists a neighborhood $U_{\lambda_0}$ of $\lambda_0$ such that for all $\lambda\in U_{\lambda_0}$ one of the following holds:
\begin{list}{(\arabic{prop}):}
{\usecounter{prop}
\lindent}
\item\label{samesign}$\rw'(\Lambda,\lambda,r_c(\Lambda,\lambda))=\pm\infty$,
\item\label{sameexit}$\rw(\Lambda,\lambda,r_c(\Lambda,\lambda))=\pm1$, or
\item\label{psdo}$(A(\Lambda,\lambda,r),\rw(\Lambda,\lambda,r))$ is noncompact.
\end{list}
\end{Lma}
\noindent
\textbf{Proof}:  From \textit{Fact}~\ref{reflect}, we may assume, without loss of generality, that
\[\rw'(\Lambda,\lambda_0,r_c(\Lambda,\lambda_0))=-\infty.\]
Because for $\lambda$ near $\lambda_0$, $\lim_{r\nearrow r_c(\Lambda,\lambda)}\rw'(r)$ exists, the only alternatives to Cases (\ref{samesign}), (\ref{sameexit}), and (\ref{psdo}) are the following:
\begin{list}{(\alph{prop}):}
{\usecounter{prop}
\lindent}
\item $\rw'(\Lambda,\lambda ,r_c(\Lambda,\lambda)=+\infty$ or 
\item$\rw(\Lambda,\lambda,r_c(\Lambda,\lambda))=1$.
\end{list}
We show that the assumption that either (a) or (b) holds for $\lambda$ sufficiently close to $\lambda_0$ leads to a contradiction.   Assuming that $r_c(\Lambda,\lambda)$ is a continuous function of $\lambda$, we prove that
\begin{list}{(\roman{prop}):}
{\usecounter{prop}
\lindent}
\item\label{rcinc}for $\lambda$ sufficiently close to $\lambda_0$, $r_c(\Lambda,\lambda)> r_c(\Lambda,\lambda_0)$ and
\item\label{rcdec} for $\lambda$ sufficiently close to $\lambda_0$, $r_c(\Lambda,\lambda)< r_c(\Lambda,\lambda_0)$.
\end{list}
Both (i) and (ii) cannot hold, so we will have the desired contradiction.

To prove (i), we choose arbitrary $M>0$.  Then,  there exist $\epsilon$ such that $\rw'(\Lambda,\lambda_0,r)<-2M$ whenever $r$ is within $\epsilon$ of $r_c(\Lambda,\lambda_0)$.  The continuity of $r_c$ implies the existence of $\delta>0$ such that $r_c(\Lambda,\lambda)>r_c(\Lambda,\lambda_0)-\epsilon$ whenever $\lambda$ is within $\delta$ of $\lambda_0$.  Continuous dependence on parameters ensures (choosing $\delta$ smaller if necessary) that $\rw'(\Lambda,\lambda,r)<-M$ whenever $\lambda$ is within $\delta$ of $\lambda_0$ and $r\in (r_c(\Lambda,\lambda_0)-\epsilon,\min\{r_c(\Lambda,\lambda_0),r_c(\lambda,\lambda)\})$; i.e., $r_c(\Lambda,\lambda)>r_c(\Lambda,\lambda_0)$ when either (a) or (b) holds; for when either holds, $\rw'(r_c(\Lambda,\lambda))>0$.  This proves (i).

To prove (ii), we note that Lemma~\ref{rcbig} states that $r_c(\Lambda,\lambda_0)>1/\sqrt{\Lambda}$.  Equation~(\ref{phidef}) then implies that there exist $c>0$ and $\tilde r<r_c(\Lambda,\lambda_0)$ such that any $\lambda$ that satisfies $r_c(\Lambda,\lambda)>\tilde r$ also satisfies $\Phi(\Lambda,\lambda,r)<-c$ whenever $r\in(\tilde r,r_c(\Lambda,\lambda))$ .  Also, by continuous dependence on parameters and the fact $\rw'(\Lambda,\lambda_0,r_c(\Lambda,\lambda_0))=-\infty$, for such $\lambda$, there exist $\hat r\in(\tilde r,r_c(\Lambda,\lambda,r_c(\Lambda,\lambda))$ such that $\rw'(\Lambda,\lambda,\hat r)<-2/(\hat r c)$ provided that $\lambda$ is also sufficiently close to $\lambda_0$.  Now, if for any of these $\lambda$ either (a) or (b) holds, then there exist $s\in(\hat r,r_c(\Lambda,\lambda))$ that satisfy $\rw^2(s)<1$, $\rw'(s)<-1/(\hat r c)$ and $\rw''(s)>0$. Thus,
\[[r^2A\rw''+r\Phi\rw'+\rw(1-\rw^2)]_{r=s}<0.\]
But this contradicts equation~(\ref{weq}).  We conclude that, for $\lambda$ sufficiently close to $\lambda_0$, $r_c(\lambda,\lambda)<r_c(\Lambda,\lambda_0)$.\hfill $\blacksquare$  
\begin{thm}\label{ppl}
For each positive integer $N$, there exist $\Lambda_N$ such that for each fixed $\Lambda\in(0,\Lambda_N)$ there exist $\{\lambda_n(\Lambda)\}_{n=1}^N$ such that the solution \\
$(A(\Lambda,\lambda_n(\Lambda),r),\rw(\Lambda,\lambda_n(\Lambda),r))$ is noncompact and $\rw(\Lambda,\lambda_n(\Lambda),r)$ has at least $n$ zeros between $r=0$ and $r=r_c(\Lambda,\lambda_n)$.
\end{thm}
\noindent
\textbf{Proof}:  We recall from \textit{Fact}~\ref{krusk} that it suffices to find solutions in $\mathcal{K}$ for which $\lp_{r\nearrow r_c}{\rw'}^2(r)<\infty$.  To this end, using PL~\ref{54} and continuous dependence on parameters, we can find (for sufficiently small fixed $\Lambda$) $\tilde\lambda_n$ and $\tilde\mu_n$ such that
\begin{list}{(\Roman{prop}):}
{\usecounter{prop}
\lindent}
\item $(A(\Lambda,\tilde\lambda_n,r_c(\Lambda,\tilde\lambda_n)),\rw(\Lambda,\tilde\lambda_n,r_c(\Lambda,\tilde\lambda_n)))=(A^-_c,-1)$ and \\
$(A(\Lambda,\tilde \mu_n,r_c(\Lambda,\tilde\lambda_n)),\rw(\lambda,\tilde\mu_n,r_c(\Lambda,\tilde\mu_n)))=(A^+_c,1)$ where $A^-_c$ and $A^+_c$ are both strictly positive, and
\item $\rw(\Lambda,\tilde\lambda_n,r)$ and $\rw(\Lambda,\tilde\mu_n,r)$ both have at least $n$ zeros before their respective crash points.
\end{list}
Condition (II) follows from the Implicit Function Theorem and \textit{Fact}~\ref{wconst}.
Without loss of generality, we assume that $\tilde\lambda_n<\tilde \mu_n$.  If this is not the case, we simply interchange their roles in what follows.  Next, we define
\begin{list}{}
{\lindent}
\item $\hat\lambda_n=\sup\{\tilde\lambda_n<\tilde \mu_n\mathrm{\;that\; satisfy\; (I)\; and\; (II)}\}$ and
\item  $\hat\mu_n=\inf\{\tilde\mu_n>\hat\lambda_n\mathrm{\;that\;satisfy\;(I)\;and\;(II)}\}$.
\end{list}

Clearly $\hat\lambda\le \hat \mu$.  It follows from Lemma~\ref{wpinfopen}, Lemma~\ref{Arc+} and the definitions of $\hat\lambda$ and $\hat\mu$ that the inequality is strict; i.e., $\hat\lambda<\hat\mu$ and that $\{\Lambda\}\times [\hat\lambda,\hat \mu]\in\mathcal{K}_c$.  From Lemma~\ref{limits} it also follows that for all $\lambda\in [\hat\lambda,\hat\mu]$, either ${\rw'}^2(r_c(\Lambda,\lambda))=\infty$ or $(A(\Lambda,\lambda,r),\rw(\Lambda,\lambda,r))$ is noncompact.

We now define
\begin{list}{}
{\lindent}
\item $E^+=\{\lambda\in[\hat\lambda_n,\hat\mu_{n}]:
\rw'(\Lambda,\lambda,r_c(\Lambda,\lambda))=\infty\}$ and
\item $E^-=\{\lambda\in[\hat\lambda_n,\hat\mu_{n}]:
\rw'(\Lambda,\lambda,r_c(\Lambda,\lambda))=-\infty\}.$
\end{list}
Lemma~\ref{wpinfopen} implies for each $\lambda\in E^+$ the existence of an open set $U_\lambda$ containing $\lambda$ and such that $U_\lambda\cap E^-$ is empty.  Similarly, for each $\lambda\in E^-$, there exists an open set $U_\lambda$ containing $\lambda$ such that $U_\lambda\cap E^+$ is empty.  Clearly, $U^+=\bigcup_{\lambda\in E^+}U_\lambda$ and $U^-=\bigcup_{\lambda\in E^-}U_\lambda$ are open sets.  Also, $U^+\cap E^-$ and $U^-\cap E^+$ are both empty.   Now, either $U^+$ and $U^-$ are disjoint or they have nonempty intersection. If they are disjoint, then because $[\hat\lambda,\hat\mu]$ is connected, there exists at least one $\lambda_p$ such that $\lambda_p\in[\hat\lambda,\hat\mu]$ but $\lambda_p\notin E^-\cup E^+$.  If they are not disjoint, then there exists a $\lambda_p\in U^+\cap U^-$ and again, $\lambda_p\not\in E^+\cup E^-$.  $(A(\Lambda,\lambda_p,r),\rw(\Lambda,\lambda_p,r))$ is, therefore, noncompact.  \hfill $\blacksquare$

\ms{Proofs of Technical Lemmas}
\label{tech}
$\;$

In this section we establish the claims made in  Section~\ref{Pseudo} that were used to prove Theorem~\ref{ppl}.  The main goals are to establish the continuity of $r_c$ as a function of $\lambda$ for fixed $\Lambda$ and to establish limits on $A'$, $\rw$, and $\rw'$ as $r$ approaches $r_c$ .  For technical reasons, the possibilities are broken down as follows:
\begin{list}{(\arabic{prop}):}
{\usecounter{prop}
\lindent}
\item\label{reexit} $r_c=r_e$; i.e., an orbit leaves $\Gamma$ with $A>0$ and $\rw^2=1$,
\item\label{rcsmall}$A(r_c)=0$ and $r_c\le\sqrt{2}$,
\item\label{rcmid}$A(r_c)=0$ and $\sqrt{2}\le r_c\le 1/\sqrt{\Lambda}$, and
\item\label{rcbenough}$A(r_c)=0$ and $1/\sqrt{\Lambda}<r_c\le \sqrt{3/\Lambda}$.
\end{list}
We have already proved the continuity of $r_c$ in Case~\ref{reexit}.  (See Lemma~\ref{Arc+}.)  We will exclude Case~\ref{rcsmall}.  The reason for choosing $R=\sqrt{2}$ is that with this choice, in either of Case~\ref{rcmid} and Case~\ref{rcbenough}, we can establish limits on $\rw'$ and $A'$ as $r\nearrow r_c$.  Case~\ref{rcmid} will then be shown to be impossible.  Finally, we will prove the continuity of $r_c$ in Case~\ref{rcbenough}.
\subsection{Limits of $\rw$ and $A'$}
\label{varlims}
$\;$

We begin by eliminating Case~\ref{rcsmall} and establishing the limits on $A'$, $\rw$, and $\rw'$ if $r_c$ is in the remaining region.
\begin{Lma}\label{compact}
Let $R>1$ be arbitrary and define
\[\Gamma_R=\Gamma\cap \{r,A,\rw,\rw':0<r<R\}.\] 
There exist $\bar\lambda(0)$, $\epsilon>0$ and $\tilde\Lambda$ such that for any $\Lambda\in[0,\tilde\Lambda]$ and any
$\lambda\in[0,\bar \lambda(0))$, the solution $A(\Lambda,\lambda,r),\;\rw(\Lambda,\lambda,r)$ exits $\Gamma_R$ at some $r_e \le R$ and
$A(r,\Lambda,\lambda)>\epsilon$ in the interval $[0,r_e]$.
\end{Lma}
\noindent
\textbf{Proof}:   We define $\bar\lambda(0)=\lim_{n\nearrow \infty} \lambda_n(0)$
where $\lambda_n(0)$ is the value of $\lambda$ that produces the $\mathrm{n^{th}}$ particlelike solution.  For any $\lambda<\bar\lambda(0)$, necessarily $A(0,\lambda,r)>0$ for all $r\in 
[0,r_e(0,\lambda)]$ (\cite{jS93} Theorem
3.1).
Therefore, for any $\bar\lambda<\lambda(0)$ and any $\lambda<\bar\lambda$, there are two possibilities:
\begin{list}{(\Alph{prop}):}
{\usecounter{prop}
\lindent}
\item\label{appl}$A(0,\lambda,r)$ is a particlelike solution, or
\item\label{re=1}$\rw^2(0,\lambda,r_e)=1$ for some $r_e>0$.
\end{list}

In Case~A, $A>0$ for all $r>0$.  Continuous dependence on parameters ensures that for any $R>0$, there exists a $(\Lambda,\lambda)$-neighborhood $U_\lambda$ of $(0,\lambda)$ such that 
$A(\Lambda,\lambda,r)>0$ whenever $(\Lambda,\lambda)\in U_\lambda$ and $0\le r\le R$.

In Case~B, Lemma~\ref{Arc+} also implies the existence of a $(\Lambda,\lambda)$-neighborhood $U_\lambda$ such that  $\rw^2(\Lambda,\lambda,r_e(\Lambda,\lambda))=1$ and $A>0$ throughout the interval $[0,r_e(\Lambda,\lambda))$ whenever $(\Lambda,\lambda)\in U_\lambda$.  ($r_e$ is the point at which the solution exits $\Gamma$.)

Thus, we have for each $\lambda$, a neighborhood $U_\lambda$ such that whenever $(\Lambda,\lambda)\in U_\lambda$, the solution $(A(\Lambda,\lambda,r),\rw(\Lambda,\lambda,r))$ exits $\Gamma_R$ at some $r^R_e(\Lambda,\lambda)\le R$ and $A(\Lambda,\lambda,r))>0$ throughout the interval $[0,r^R_e]$.  The result now follows because the interval $[0,\bar\lambda]$ is compact.  \hfill $\blacksquare$\\ 

Throughout the rest of this paper, we fix $R=\sqrt{2}$ and unless stated otherwise, assume solutions lie in the set $\mathcal{K}$ defined by $\tilde\Lambda$ and $\lambda_0$ of Lemma~\ref{compact}.  As an obvious consequence of Lemma~\ref{compact} we have, for such solutions, the following:
\begin{Lma}\label{bigenough}
Suppose $A(r_c)=0$.  Then $r_c>\sqrt{2}$.
\end{Lma}
\indent
The next Lemma is crucial to establishing the continuity of $r_c$.
\begin{Lma}\label{limits}
Suppose $\lp_{r\nearrow r_c}\rw^2(r)<\infty$.  Then $A{\rw'}^2$, $\Phi$, $A'$, and $\rw$ all have finite limits as $r\nearrow r_c$.
\end{Lma}
\noindent
\textbf{Proof}:  We first note that from equation~(\ref{Aeq}) it is clear that the existence and finiteness of any two of the 
limits of $A{\rw'}^2$, $A'$ and $\Phi$ implies the existence and finiteness of the third.  Also $\lim_{r\nearrow
r_c}\Phi(r)$ exists if and only if $\lim_{r\nearrow r_c}\rw(r)$ exists.

In Lemma~\ref{A+limits} we already proved the result in the case where $A(r_c)>0$.  Thus, we may assume $A(r_c)=0$.  We define 
\begin{equation}\label{z}
z(r)=\frac{\Phi}{r}+\frac{2A{\rw'}^2}{r}.
\end{equation}
A simple calculation using equations~(\ref{Aeq}) and (\ref{weq}) yields
\begin{equation}\label{zeq}
r^2z'+2r{\rw'}^2z+2 (1-A-2\frac{(1-\rw^2)^2}{r^2})=0.
\end{equation}

Because $r_c^2>2$, the last term on the left side of equation~(\ref{zeq}) is strictly positive whenever $r$ is sufficiently close to $r_c$.  Thus, near $r_c$, we cannot have $z(r)=z'(r)=0$.  Also, $z'<0$ whenever $z=0$.  Therefore, 
$z$ has only one sign near $r_c$. There are now two cases to consider:
\begin{list}{(\Alph{prop}):}
{\usecounter{prop}
\lindent}
\item\label{z+} $z >0$ near $r_c$ and
\item\label{z-} $z <0$ near $r_c$.
\end{list} 
In either case, since the last term on the left side of equation~(\ref{zeq}) is bounded, $z'$ is bounded from one side or the other; i.e., $\lim_{r\nearrow r_c}z(r)$ exists.\\ \\
\textit{Case~A}.  $z>0$ near $r_c$.  Equation~(\ref{zeq}) implies $z'<0$ near $r_c$.  Therefore 
$\lim_{r\nearrow r_c}z(r)$ is finite.  We consider the two subcases: 
\begin{list}{(A\roman{prop}):} 
{\usecounter{prop}
\lindent}
\item $\lim_{r\nearrow r_c}z >0$ and
\item $\lim_{r\nearrow r_c}z=0$.
\end{list}
\textit{Case~A}i.  We prove that the assumption that $A{\rw'}^2$ has no limit leads to a contradiction.  For under this assumption, there exists a sequence
$\{r_n\}$ such that $(A{\rw'}^2)'(r_n)=0$ and $A{\rw'}^2(r_n)>\epsilon$; i.e., 
${\rw'}^2(r_n)\rightarrow \infty$.  Evaluating equation~(\ref{feq}) at $r_n$
gives 
\[\rw'[r^2{\rw'}z + 2\rw(1-\rw^2)]_{r=r_n}=0.\]
As $n\nearrow \infty$, the first term in parentheses dominates the second term since the second term remains bounded while the first term is unbounded; i.e., for sufficiently large $n$, the expression on the left cannot equal 0.  This proves that $A{\rw'}^2$ has a limit.

Since $z$ also has a limit,
$\Phi$ and $\rw$ must also have limits.  Furthermore, $\lim_{r\nearrow r_c}\Phi(r)$ is finite because $\rw$ is bounded.  Thus
$\lim_{r\nearrow r_c}A{\rw'}^2$ is also finite.\\ \\
\textit{Cases~A}i and~\textit{B}:  By hypothesis, $\Phi$ is bounded near $r_c$.  Thus, since $z$ is bounded from above and $A{\rw'}^2\ge 0$, ${A\rw'}^2$ is also bounded.  It follows that 
\[-\infty<\lim_{r\nearrow r_c}z(r)\le 0.\]  

We now define 
\begin{equation}\label{y}
y=\rw\rw'(1-\rw^2).
\end{equation}
A straighforward calculation using equations~(\ref{Aeq}) and (\ref{weq}) yields
\begin{equation}\label{yeq}
r^2Ay'+r\Phi y+\rw^2(1-\rw^2)^2-r^2A{\rw'}^2(1-3\rw^2)=0.
\end{equation}  
$z\le 0$ and $A{\rw'}^2\ge0$ imply that $\lp_{r\nearrow r_c}\Phi(r) \le 0$.

We prove that the assumption $\lim_{r\nearrow
r_c}\Phi(r)$ does not exist leads to a contradiction.  Indeed, under this assumption,  for some $\epsilon >0$ and any $M>0$, there exist $r^M_0$ and $r^M_1$,
$r^M_0<r^M_1$ such that $(r\Phi)\le -\epsilon$ in the interval
$[r^M_0,r^M_1]$, $(r\Phi)(r^M_0)=(r\Phi)(r^M_1)=-\epsilon$, and $(r\Phi)'(r^M_0)
<-M$ (See Figure~2).  We denote by
$r^M_m$ the point on $(r^M_0,r^M_1)$ where $r\Phi$ is minimized.\\ \\
\begin{center}
Figure~2.
\end{center}
\nopagebreak
\setlength{\unitlength}{0.7mm}
\begin{picture}(0,100)(-60,20)\label{zlimit}
\put (-30,85) {\vector(1,0){130}}
\put (-30,70) {\line(1,0){130}}
\put (-30,30) {\vector(0,1){65}}
\put (-33,96.5){$r\Phi$}
\put (101,84){$r$}
\put (-33.5,83.5){$0$}
\put (-37.5,68.5) {$-\epsilon$}
\put (16,87) {$r_0^M$}
\put (25.9,87) {$r_m^M$}
\put (34.9,87) {$r_1^M$}
\put (37.9,70) {\dashbox (0,15)}
\put (18.6,70) {\dashbox (0,15)}
\put (28.25,27) {\dashbox (0,58)}
\qbezier (18,75) (28,-20)(38,73)
\qbezier (38,73) (41,85) (42,30)
\qbezier (5.18,30) (17.31,100) (18,73)
\put (43,15){\dashbox(0,80)}
\put (44,20){$r=r_c$}
\end{picture}
$\;$\\ \\
Since the last two terms on the left side of equation~(\ref{yeq}) are bounded, there exists a positive $B_1$ such that
\begin{equation}\label{yineq}
r^2Ay'<-(r\Phi)y+B_1.
\end{equation}
Now, a simple calulation using equations~(\ref{Aeq}) and~(\ref{phidef}) yields
\begin{equation}\label{rphider}
(r\Phi)'-2A{\rw'}^2-\frac{2(1-\rw^2)^2}{r^2}+2\Lambda r^2-\frac{4y}{r}=0.
\end{equation}
The middle three terms on the left side of equation (\ref{rphider}) are also bounded; i.e., there exist $B_2>0$ such that
\begin{equation}\label{rpineq}
y<\frac{r(r\Phi)'}{4}+B_2.
\end{equation}  
Inequality~(\ref{rpineq}) allows us to choose $M$ sufficiently large so that 
\begin{equation}\label{yinit}
y(r_0^M)<\frac{-B_1}{\epsilon}.
\end{equation}
Substituting inequality~(\ref{yinit}) into inequality~(\ref{yineq}) gives
\begin{equation}\label{y'-}
y'(r_0^M)<0.
\end{equation}
We claim that
\begin{equation}\label{y'-unif}
y'(r)<0\;\;\;\mathrm{for\;all}\;r\in [r_0^M,r_1^M].
\end{equation}
Indeed, $(r\Phi)<-\epsilon$ in the open interval $(r_0^M,r_1^M)$.  If there exists an $s$ in this interval that satisfies $y'(s)=0$, we take $s$ to be as small as possible.  Clearly, $y(s)<y(r_0^M)$.  However, inequalities~(\ref{yineq}) and (\ref{yinit}) imply that
\[y(s)=\frac{B_1}{s\Phi(s)}>\frac{B_1}{r^M_0\Phi(r_0^M)}=\frac{-B_1}{\epsilon}>y(r_0^M).\]
Thus, there can be no such $s$.  Inequality~(\ref{y'-unif}) follows.

Inequality~(\ref{rpineq}) now yields, for $r_0^M$ sufficiently close to $r_c$ and all $r\in(r_0^M,r_1^M]$, 
\begin{eqnarray}\label{p-}
(r\Phi)'(r)&<&\frac{4(y(r)-B_2)}{r}<\frac{4(y(r_0^M)-B_2)}{r}\nonumber\\
 &<&\frac{4(y(r_0^M)-B_2)}{r_0^M}+\frac{M}{2}\nonumber\\
&<&(r\Phi)'(r_0^M)+\frac{M}{2}<0.
\end{eqnarray}
(The third inequality holds because under the assumption that $\Phi(r)$ has no limit as $r\nearrow r_c$, $r_0^M$ and $r_1^M$ can be made arbitrarily close.)  In particular, inequality~(\ref{p-}), evalutated at $r_m^M$ gives $r_m^M<0$ which is impossible.  This proves $\Phi$ has a limit.  Clearly, it is finite.  Since $\lim_{r\nearrow r_c}z(r)$ exists and is finite, the result follows. \hfill $\blacksquare$
\nopagebreak
\par
\indent
\subsection{Uniform bound on $r_c$}
\label{bound}
$\;$

To establish the existence of $\lim_{r\nearrow r_c}\rw'(r)$ and to prove the continuity of $r_c$ we must eliminate the possibility that $r_c\in J_\Lambda$.  The desired result is the following:
\begin{Lma}\label{rcbig}
There exist $\bar\Lambda$ such that for all $\Lambda\in(0,\bar\Lambda)$, for all $\lambda$,
\[r_c(\Lambda,\lambda)>1/\sqrt{\Lambda}\]
whenever $A(r_c(\Lambda,\lambda))=0$, $\lp_{r\nearrow r_c(\Lambda,\lambda)}\rw^2\le 1$,and 
$\lp_{r\nearrow r_c(\Lambda,\lambda)}{\rw'}^2(r)=\infty$. 
\end{Lma}
\noindent
\textbf{Proof}:  Without loss of generality, we assume there is a sequence $\{r_n\}\rightarrow r_c$ such that $\rw'(r_n)>(1-1/\sqrt{3})/\sqrt{6}$ and $\rw''(r_n)>0$.  Equation~(\ref{weq}) and Lemma~\ref{limits} now yield
\[\lim_{r\nearrow r_c}\Phi(r)\le 0.\]

Multiplying equation~(\ref{phidef}) by $r^2$, evaluating at $r_c$, and solving as a quadratic in $r_c^2$ either
\begin{equation}\label{philbound}
0<r_c^2<\frac{1-\sqrt{1-4\Lambda}}{2\Lambda}\;\;\;\mathrm{or}
\end{equation}
\begin{equation}\label{phiubound}
\frac{1+\sqrt{1-4\Lambda}}{2\Lambda}\le r_c^2.
\end{equation}
Since $\sqrt{1-4\Lambda}=1-2\Lambda+\circ(\Lambda^2)$, any $r_c$ that satisfies equation~(\ref{philbound}) is less than $\sqrt{2}$ provided $\Lambda$ is sufficiently small.  Therefore, because of Lemma~\ref{bigenough}, equation~(\ref{philbound}) can be ignored.

We also consider only $\Lambda$ sufficiently small so that
\[\sqrt{(1+\sqrt{1-4\Lambda})/(2\Lambda)}>1/\sqrt{\Lambda}-2\]
and prove that, choosing $\Lambda$ smaller if necessary, whenever $A(r_c)=0$, $\rw^2(r_c)\le 1$, and $r_c\in[1/\sqrt{\Lambda}-2,1/\sqrt{\Lambda}]$, there exists an $\bar r>0$ such that either $A(\bar r)=0$ or $\rw^2(\bar r)=1$.  Thus, such a solution cannot be in the family of \textit{Fact}~\ref{family}.  This will complete the proof.

We simplify notation by setting
\[\;a_\Lambda=\frac{2}{3\sqrt{\Lambda}}-3\sqrt{3},\;b_\Lambda=\frac{2}{3\sqrt{\Lambda}},\;\mathrm{and}\;c_\Lambda=\frac{1}{\sqrt{\Lambda}}-3.\]
Lemma~\ref{A<f} states that for any of the one parameter solutions of \textit{Fact}~\ref{family}, $A{\rw'}^2(r)<A$ throughout the interval $[b_\Lambda,c_\Lambda]$ provided $\Lambda$ is sufficiently small.  Lemma~\ref{rfbound} states that there exist $K>0$ such that under the same hypotheses, $A<K/r$ throughout the same interval.  These lemmas, equations~(\ref{mu}) and~(\ref{mu'}) yield
\begin{equation}\label{ArK}
\mu'<2\tilde K/r
\end{equation}
throughout the interval $[b_\Lambda,c_\Lambda]$ for some $\tilde K>K$.  Also, Lemma~\ref{fbound} gives
\begin{equation}\label{ubound2}
\mu'<2\tilde M
\end{equation}
throughout the interval $[c_\Lambda,r_c]$ for some positive constant $\tilde M$.

On one hand, integrating inequalities~(\ref{ArK}) and ~(\ref{ubound2}) gives
\begin{eqnarray}\label{muest}
\mu(r_c)-\mu(b_\Lambda) & = & \int_{s=b_\Lambda}^{c_\Lambda}\mu'(s)\;\rd s+\int_{s=c_\Lambda}^{r_c} \mu'(s)\;\rd s\nonumber\\
 & < & \int_{s=b_\Lambda}^{c_\Lambda}2\tilde K/s\;\rd s+\int_{s=c_\Lambda}^{r_c}2\tilde M\; \rd s\nonumber\\
& \le& 2\tilde K\ln(\frac{c_\Lambda}{b_\Lambda})+6\tilde M\nonumber\\
 &=&2\tilde K\ln(3/2-9\sqrt{\Lambda}/2)+6\tilde M.
\end{eqnarray}
It is clear that for sufficiently small $\Lambda$,
\begin{equation}\label{mubound}
\mu(r_c)-\mu(b_\Lambda)<L
\end{equation}
where $L$ is any number satisfying 
\[L>2
\tilde K\ln(3/2)+6\tilde M.\]

On the other hand, we consider also 
\begin{equation}\label{xdef}
h(r)=\mu(r)+rA(r)=r(1-\frac{\Lambda r^2}{3}).
\end{equation}
$h'(r)=1-\Lambda r^2>0$ for all $r\in (0,r_c)$.  Therefore,
\begin{equation}\label{hrc}
h(r_c)-h(b_\Lambda)>h(c_\Lambda)-h(b_\Lambda).
\end{equation}
A simple calculation yields
\begin{equation}\label{delh}
h(c_\Lambda)-h(b_\Lambda)=\frac{8}{81\sqrt{\Lambda}}+\circ (\Lambda^0).
\end{equation}
Inequality~(\ref{hrc}) and equation~(\ref{delh}) together imply that  
\begin{equation}\label{dh}
h(r_c)-h(b_\Lambda)>L
\end{equation}
for $\Lambda$ sufficiently small.  Now, $\mu(r_c)=h(r_c)$.  So 
comparing inequality~(\ref{dh}) to inequality~(\ref{mubound}) gives
\begin{equation}\label{hcross}
\mu(b_\Lambda)>h(b_\Lambda)
\end{equation}
if $\Lambda$ is sufficiently small.  Also because $\mu(r_c)=h(r_c)$, either $\mu(r)>h(r)$ for all $r\in(b_\Lambda,r_c)$ or there exists an $\bar r\in(b_\Lambda,r_c)$ such that $\mu(\bar r)=h(\bar r)$.  In the former case, equation~(\ref{xdef}) gives $A<0$ in $(b_\Lambda,r_c)$.  We therefore rule this case out.  In the latter case, equation~(\ref{xdef}) gives $A(\bar r)=0$.  This completes the proof assuming Lemmas~\ref{fbound}, \ref{A<f} and \ref{rfbound}.  \hfill $\blacksquare$
\begin{Lma}\label{fbound}
There exist $M>0$ such that for all $(\Lambda,\lambda)\in \mathcal{K}_c$,\\
$A{\rw'}^2(\Lambda,\lambda,r)<M$ for all $r\in [0,r_c(\Lambda,\lambda)]$.
\end{Lma}
\noindent
\textbf{Proof}:  Lemma~\ref{compact} gives, for all solutions in $\mathcal{K}$, an $r_e\le \sqrt{2}$ such that $(A,\rw)$ exits $\Gamma_{\sqrt{2}}$ at $r_e$ and $A(r_e)>0$.  We define
\[\rho(\Lambda,\lambda)=\min\{1,r_e(\Lambda,\lambda)\}\;\;\;\mathrm{and}\]
\[\Delta=\{(\Lambda,\lambda,r):(\Lambda,\lambda)\in \mathcal{K}\; \mathrm{and}\;0\le r\le \rho(\Lambda,\lambda)\}.\]
Standard results and Lemma~\ref{a=0} imply that any solution $(A(\Lambda,\lambda,r),\rw(\Lambda,\lambda,r))$ in $\mathcal{K}$ can be extended beyond $\rho(\Lambda,\lambda)$.  It follows from continuous dependence of solutions on parameters that $\Delta$ is a closed subset of $\RR^3$.  Being bounded, it is also compact.  Therefore, there exists an $M_1$ such that
\begin{equation}\label{r<1fbound}
A{\rw'}^2(r)<M_1\;\;\;\mathrm{for\; all}\; (\Lambda,\lambda,r)\in\Delta.
\end{equation}
We have $A{\rw'}^2(r)<M_1$ for $r\in(0,r_c)$ whenever $\rho<1$.  To find a bound when $\rho=1$ we define $M=\max\{M_1,2+2/\sqrt{27}\}$ and recall equation~(\ref{feq}),
\[\;\;\;\;\;\;\;\;\;\;\;\;\;r^2(A{\rw'}^2)'+\rw'[r\rw'(\Phi+2A{\rw'}^2)+2\rw(1-\rw^2)]=0.\;\;\;\;\;\;\;\;\;\;\;\;\;\;\;\;\;(\ref{feq})\]
In the interval $[1,r_c]$ $0<A<1$, $|\rw(1-\rw^2)|<2/(3\sqrt{3})$, and, because of Theorem~\ref{Acrashes}, $\Phi>-4$.  Also, for all $\tilde r$ in this interval, whenever $A{\rw'}^2(\tilde r)>M$, ${\rw'}^2(\tilde r)>1$.  This and equation~(\ref{feq}) imply that in the interval $(1,r_c)$,
\begin{equation}\label{f'-}
(A{\rw'}^2)'(\tilde r)<0\;\mathrm{whenever}\;A{\rw'}^2(\tilde r)>M.
\end{equation}
Inequalities~(\ref{r<1fbound}) and (\ref{f'-}) imply that $A{\rw'}^2$ cannot exceed $M$ in the interval $(1,r_c]$.  (\ref{r<1fbound}) also implies that $A{\rw'}^2$ cannot exceed $M\;(>M_1)$ in the interval $[0,1]$.  The result follows. \hfill $\blacksquare$    

We now improve on this bound in the interval $[b_\Lambda,c_\Lambda]$.
\begin{Lma}\label{A<f}
For $\Lambda$ sufficiently small, any solution $(A,\rw)$ in $\mathcal{K}_c$ that satisfies $A(r_c)=0$ with $r_c>c_\Lambda+1$
also satisfies ${\rw'}^2<1$ throughout the interval $[b_\Lambda,c_\Lambda]$.
\end{Lma}
\noindent
\textbf{Proof}:  We first prove that
\begin{equation}\label{P+A}
\Phi+A>\frac{1}{r}\;\;\;\mathrm{for\;all\;}r\in[a_\Lambda,c_\Lambda]
\end{equation}
whenever $\Lambda$ is sufficiently small, $(\Lambda,\lambda)\in \mathcal{K}_c$, and $r_c>c_\Lambda+1$.  To this end, we consider $\Lambda$ sufficiently small so that $a_\Lambda>0$ and examine the function
\begin{equation}\label{psi}
\psi(r)=1-\frac{1}{r^2}-\Lambda r^2-\frac{1}{r}.
\end{equation}
We will prove that $\psi(r)>0$ for all $r\in(a_\Lambda,c_\Lambda)$.  The result will follow.

Now,
\begin{equation}\label{psiL}
(r\psi)(\frac{1}{\sqrt{\Lambda}})=r\psi(c_\Lambda+3)=-1-\sqrt{\Lambda}.
\end{equation}
Also, 
\begin{equation}\label{rpsi'}
(r\psi)'(r)=1+\frac{1}{r^2}-3\Lambda r^2.
\end{equation}
From equation~(\ref{rpsi'}), it is clear that 
\begin{equation}\label{psiunif}
(r\psi')(r)\rightarrow -2\;\;\;\mathrm{uniformly\;in}\;[c_\Lambda,r_c]\;\mathrm{as}\;\Lambda\searrow 0.
\end{equation}
Equation~(\ref{psiL}) and~(\ref{rpsi'}) imply that, for sufficiently small $\Lambda$,
\begin{equation}\label{psi+}
\psi(c_\Lambda)>0.
\end{equation}
Simple calculations give
\begin{equation}\label{psider}
\psi'(r)=\frac{2}{r^3}-2\Lambda r +\frac{1}{r^2}
\end{equation}
and
\begin{equation}\label{psidd}
\psi''(r)=-\frac{6}{r^4}-2\Lambda-\frac{2}{r^3}.
\end{equation}
From equations~(\ref{psider}) and~(\ref{psidd}), it follows readily that
\begin{equation}\label{psi'L}
r^4\psi'(a_\Lambda)=-\frac{32}{81\Lambda}+\circ(\frac{1}{\sqrt{\Lambda}})
\end{equation}
and
\begin{equation}\label{psidd+}
\psi''(r)<0\;\;\;\mathrm{for\;all}\;r>0.
\end{equation}
For $\Lambda$ sufficiently small, the right side of equation~(\ref{psi'L}) is negative.  This and equation~(\ref{psidd+}) imply that 
\begin{equation}\label{psi'-unif}
\psi'(r)<0\;\;\;\mathrm{for\;all}\;r\in(a_\Lambda,c_\Lambda).
\end{equation}
Equations~(\ref{psi+}) and (\ref{psi'-unif}) establish that
\begin{equation}\label{psi+u}
\psi(r)>0\;\;\;\mathrm{for\;all}\;r\in[a_\Lambda,c_\Lambda].
\end{equation}
Finally,
\[\psi=\Phi+A+[\frac{(1-\rw^2)^2}{r^2}-\frac{1}{r^2}]-\frac{1}{r}\le\Phi+A-\frac{1}{r}.\]
Inequality~(\ref{P+A}) follows.

We now define the set
\begin{equation}\label{Wset}
W=\{r\in(a_\Lambda,c_\Lambda):|\rw'(r)|\le \frac{2}{3\sqrt{3}}\}.
\end{equation}
$W$ is not empty.  In fact,
\begin{equation}\label{Wne}
W\cap(a_\Lambda,b_\Lambda)\;\;\;\mathrm{is\; not\;empty.}
\end{equation}
Indeed, otherwise, without loss of generality, (\textit{Fact}\ref{reflect}), we may assume that $\rw'(r)>2/(3\sqrt{3})$ for all $r\in (a_\Lambda,b_\Lambda)$.  Integrating this yields $\rw(b_\Lambda)-\rw(a_\Lambda)>2$, contradicting the assumption that $|\rw|\le 1$ in $[0,r_c]$.  This establishes (\ref{Wne}).

Next, we define
\begin{equation}\label{wsup}
\hat r=\sup\{r\in W\}.
\end{equation}
Again using \textit{Fact}~\ref{reflect}, we assume that $\rw'>2/(3\sqrt{3})$ throughout the interval $(\hat r,c_\Lambda)$.  Equations~(\ref{weq}) and (\ref{P+A}) yield
\begin{equation}\label{w+-A}
\rw''=\frac{1}{r^2A}[-r(\Phi+A)\rw'-\rw(1-\rw^2)]+\frac{\rw'}{r}<\frac{\rw'}{r}\;\mathrm{\;\;in\;}(\hat r,c_\Lambda).
\end{equation}
Integrating inequality~(\ref{w+-A}) from $\hat r$ to $r$ with the condition $\rw'(\hat r)=2/(3\sqrt{3})$ yields
\begin{equation}\label{w''integrated}
\rw'(r)<\frac{2r}{3\sqrt{3}\hat r}\;\;\;\mathrm{for\;all\;}r\in(\hat r,c_\Lambda).
\end{equation}
In particular,
\begin{equation}\label{w'good}
\rw'(r)<1\;\;\;\mathrm{for\;all\;}r\in(\hat r,\min\{c_\Lambda,3\sqrt{3}\hat r/2\}).
\end{equation}
Now,
\begin{eqnarray}\label{rc>cl}
\frac{3\sqrt{3}\hat r}{2}&>&\frac{3\sqrt{3}a_\Lambda}{2}=\frac{3\sqrt{3}}{2}[\frac{2}{3\sqrt{\Lambda}}-3\sqrt{3}]\nonumber\\
&=&\frac{\sqrt{3}}{\sqrt{\Lambda}}-\frac{27}{2}.
\end{eqnarray}
It follows easily from inequality~(\ref{rc>cl}) that for small $\Lambda$, $3\sqrt{3}\hat r/2$ exceeds $c_\Lambda$.  Substituting this fact into (\ref{w'good}) yields $\rw'(r)<1$ for all $r\in [b_\Lambda,c_\Lambda]$.\hfill $\blacksquare$
\begin{Lma}\label{rfbound}
There exists a $K$ such that for sufficiently small $\Lambda$, any solution that satisfies the hypotheses of Lemma~\ref{A<f} also satisfies $A<K/r$ in the interval $[b_\Lambda,c_\Lambda]$.
\end{Lma}
\noindent
\textbf{Proof}:  Invoking Lemma~\ref{A<f} gives from equation~(\ref{Aeq})
\begin{equation}\label{Afest}
(r^3A)'>r^2(1-\frac{1}{r^2}-\Lambda r^2)=r^2-1-\Lambda r^4>-1
\end{equation}
throughout the interval $[b_\Lambda,c_\Lambda]$.
Integrating inequality~(\ref{Afest}) from any $r\in[b_\Lambda,c_\Lambda]$ to $c_\Lambda$ yields
\begin{equation}\label{Afint}
A(r)<\frac{c_\Lambda}{r^3}-\frac{1}{r^2}+\frac{c_\Lambda^3A(c_\Lambda)}{r^3}.
\end{equation}
Now,
\begin{equation}\label{bineq}
c_\Lambda<(3/2)b_\Lambda<(3/2)r.
\end{equation}
Also, equation~(\ref{Aeq}) and Lemma~\ref{fbound} yield the existence of $M>0$ such that, for $\Lambda$ sufficiently small, 
\begin{equation}\label{ra'}
(rA)'>-2M\;\;\;\mathrm{for\;all\;}r\in[b_\Lambda,r_c].
\end{equation}
  Integrating inequality~(\ref{ra'}) from $c_\Lambda$ to $r_c$ yields 
\begin{equation}\label{a/rineq}
A(r)<6M/r\;\;\;\mathrm{for\;all\;}r\in [r_c-3,r_c].
\end{equation}
Since $r_c-3<c_\Lambda$, the result follows upon substituting (\ref{bineq}) and (\ref{a/rineq}) into (\ref{Afint}).\hfill $\blacksquare$\\ \\
\subsection{Limit of $\rw'$}
\label{rw'limit}
\begin{Lma}\label{wplimit}
If $0<\Lambda<\bar \Lambda$ ($\bar \Lambda$ as in Lemma~\ref{rcbig}), $\lim_{r\nearrow r_c}\rw'(r)$ exists.
\end{Lma}
\noindent
\textbf{Proof}: We have already proved this when $A(r_c)>0$.  Thus, we need only consider the case where $A(r_c)=0$. In view of Lemma~\ref{limits}, there are only two subcases to consider:
\begin{list}{(\arabic{prop}):}
{\usecounter{prop}
\lindent}
\item\label{phinot0} $\lim_{r\nearrow r_c}\Phi(r)\ne 0$, and
\item\label{phi0} $\lim_{r\nearrow r_c}\Phi(r)=0$.
\end{list}
\textit{Case~\ref{phinot0}}.  We suppose that there exists a sequence $\{r_n\}\nearrow r_c$ such that $\rw''(r_n)=0$ for each $n$.  Then equation~(\ref{weq}) implies
\begin{equation}\label{wpp0}
\rw'(r_n)=\frac{\rw(r_n)(\rw^2(r_n)-1)}{r_n\Phi(r_n)}.
\end{equation}
Consequently (again making use of Lemma~\ref{limits}) the right side of equation~(\ref{wpp0}) has a limit as $n\nearrow \infty$.  Since $\lim_{r\nearrow r_c}\Phi(r)\ne 0$, the result follows.\\ \\
\textit{Case~\ref{phi0}}.  There are two subcases to consider:
\begin{list}{(\ref{phi0}\alph{prop}):}
{\usecounter{prop}
\lindent}
\item $\lim_{r\nearrow r_c}\rw\ne 0$ and
\item $\lim_{r\nearrow r_c}\rw =0$.
\end{list}
\textit{Case~\ref{phi0}a}.  Equation~(\ref{weq}) implies that $\rw'$ has one sign near $r_c$.  This is because $\rw$ has only one sign near $r_c$ and whenever
$\rw'(r)=0$, $\rw''\rw(r)<0$.  Now, as in Case~\ref{phinot0}, for any sequence $\{r_n\}\nearrow r_c$ such that $\rw''(r_n)=0$ for all $n$, 
\[\rw'(r_n)=\frac{\rw(r_n)(\rw^2(r_n)-1)}{r_n\Phi(r_n)}.\] 
Clearly the right side of this equation goes to $\pm\infty$ as $n\nearrow\infty$.  Because $\rw'$ has only one sign, this must go to one or the other of $\pm\infty$.  The result follows.\\ \\  
\textit{Case~\ref{phi0}b}.  $A(r_c)=\rw(r_c)=\Phi(r_c)=0$ implies $r_c^2=\frac{1\pm
\sqrt{1-4\Lambda}}{2\Lambda}$.  But Lemmas~\ref{bigenough} and~\ref{rcbig} preclude this for small $\Lambda$.  \hfill $\blacksquare$\\
\subsection{Continuity of $r_c$}
\label{continuity}
\begin{Lma}\label{rccont}
For sufficiently small fixed $\Lambda$,  $r_c(\Lambda,\lambda)$ is a continuous function of $\lambda$.
\end{Lma}
\noindent
\textbf{Proof}:  We have already proved the continuity of $r_c$ as a function of $\lambda$ for solutions in $\bar\mathcal{K}_c$ (Lemma~\ref{Arc+} and Lemma~\ref{a=0}).  It remains to prove the result in the case $(\Lambda,\lambda)\in\mathcal{K}_c$.  From Lemma~\ref{rcbig}, we may assume that $r_c(\Lambda,\lambda)>1/\sqrt{\Lambda}$.  Since $\Lambda$ is fixed, we drop the dependence on it in what follows.

We recall the function $\mu(\lambda,r)=r(1-A(\lambda,r)-\Lambda r^2/3)$ which, for each $\lambda$, is a
nondecreasing function of $r$ (See equation~(\ref{mu'})).  Also, to simplify notation,
we define 
\[h(r)=\mu+rA=r(1-\Lambda r^2/3)\]
and 
\[\delta(\lambda,\epsilon)=h(r_c(\lambda))-h(r_c(\lambda)+\epsilon).\]
Obviously, $\mu(\lambda,r)=h(r)$ if and only if 
$A(r)=0$; i.e., only at $r_c(\lambda)$.  Furthermore, $h'(r)=1-\Lambda r^2<0$ whenever $r$ is sufficiently close to $r_c(\lambda)$.  Therefore, $\delta(\lambda,\epsilon)>0$ whenever $\epsilon>0$.  Moreover,  both $h$ and $\mu$ are continuous, $h$ is decreasing, and $\mu$ is increasing.  These facts enable us to find, for any $\epsilon>0$, $\tilde
r<r_c(\lambda)$ such that
\[\mu(\lambda,\tilde r)>h(\tilde r)-\frac{\delta(\lambda,\epsilon)}{2}.\]  
$\tilde r$ can be taken to be within $\epsilon$ of $r_c(\lambda)$.  (See Figure 3.)  Continuous dependence on 
parameters now gives $\eta>0$ such that whenever $\tilde\lambda$ is within $\eta$ of $\lambda$,
\[\mu(\tilde\lambda,\tilde r)>h(\tilde
r)-\delta.\]
Clearly, for all such $\tilde\lambda$, 
\[r_c(\tilde\lambda)>\tilde r.\]

We  now prove by contradiction that for all such $\tilde \lambda$, $r_c(\tilde\lambda)<r_c(\lambda)+\epsilon$.  If for any such $\tilde\lambda$, $r_c(\tilde\lambda)\ge r_c(\lambda)+\epsilon$, then, $\mu(\tilde\lambda,r)$ is well defined up to $r_c(\lambda)+\epsilon$ and is a continuous function of $r$ in the interval $(\tilde r,r_c(\lambda)+\epsilon)$.  Now, on one hand, 
\begin{equation}\label{h<u}
\mu(\tilde\lambda,r_c(\lambda)+\epsilon)\ge\mu(\tilde\lambda,\tilde r)>h(\tilde r)-\delta>h(r_c(\lambda))-\delta=h(r_c(\lambda)+\epsilon).
\end{equation}
On the other hand, because $A(\tilde\lambda,\tilde r)>0$, 
\begin{equation}\label{h>u}
\mu(\tilde\lambda,\tilde r)=h(\tilde r)-\tilde rA(\tilde\lambda,\tilde r)<h(\tilde r).
\end{equation}
Equations~(\ref{h<u}), (\ref{h>u}), and the Intermediate Value Theorem applied to $h(r)-\mu(\tilde\lambda,r)$ (with fixed $\tilde \lambda$) guarantee an $r_c(\lambda)\in(\tilde r,r_c(\lambda)+\epsilon)$ such
that 
\[\mu(\tilde\lambda,r_c(\lambda))=h(r_c(\tilde\lambda));\]
 i.e., $A(\tilde\lambda,r_c(\tilde\lambda))=0$.  This contradicts the assumption that $r_c(\tilde\lambda)>r_c(\lambda)+\epsilon$ and completes the proof.  \hfill $\blacksquare$ \\ \\
\begin{center}
Figure 3.
\end{center}
\begin{picture}(0,100)(-40,0)\label{uh}
\put (0,0){\vector(0,1){96}}
\put (0,0) {\vector(1,0){116}}
\put (117,-1){$r$}
\qbezier (0,0)(58,150),(116,0)
\qbezier (0,0)(40,25)(80,64.4)
\qbezier (0,0)(80,55)(100,55)
\put (80,0){\dashbox(0,39.5)}
\put (62,0){\dashbox(0,74.62)}
\put (98,0){\dashbox(0,39.5)}
\put (64,0){\dashbox(0,74.29)}
\put (79,51){$\delta$}
\put (80,50.5){\vector(0,-1){10.5}}
\put (80,55){\vector(0,1){9}}
\put(71,38){$\epsilon$}
\put(88,38){$\epsilon$}
\put(70,39.5){\vector(-1,0){8}}
\put(74,39.5){\vector(1,0){6}}
\put(87,39.5){\vector(-1,0){7}}
\put(91,39.5){\vector(1,0){7}}
\put (58,0){\dashbox(0,75)}
\put (54,-5.5){$\frac{1}{\sqrt{\Lambda}}$}
\put (77,-5.5){$ r_c$}
\put (89.6,0){\dashbox(0,52.9)}
\put (87.2,-5.5){$\tilde r_c$}
\put (63,-5.5){$\tilde r$}
\put (69,60){$\mu$}
\put (103,55){$\mu(\tilde \lambda,r)$}
\put(5,40){$h(r)$}
\put (5,-15){$\mu=\mu(\lambda,r),\;r_c=r_c(\lambda),\;\tilde r_c=r_c(\tilde\lambda)$}
\end{picture}

\bibliographystyle{plain}
\bibliography{b2}

\nocite{nS96}
\nocite{pB94}
\nocite{wR77}
\nocite{jS97}
\nocite{jS98}
\nocite{L2}
\end{document}